\documentclass[showpacs,twocolumn,aps,prl,superscriptaddress]{revtex4}
\usepackage{graphicx} % Include figure files
\usepackage{amssymb}
\begin{document}

\title{Spontaneous chirality via long-range electrostatic forces}
\author{Kevin L. Kohlstedt}
\affiliation{Department of Materials Science, Northwestern
University, Evanston, Illinois 60208}
\author{Francisco Solis}
\affiliation{Department of Integrated Natural Sciences, Arizona
State University, Glendale, Arizona 85306}
\author{Graziano Vernizzi}
\affiliation{Department of Materials Science, Northwestern
University, Evanston, Illinois 60208}
\author{Monica Olvera de la Cruz}
\affiliation{Department of Materials Science, Northwestern
University, Evanston, Illinois 60208}
\email{m-olvera@northwestern.edu}

\date{\today}

\begin{abstract}
We consider a model for periodic patterns of charges constrained
over a cylindrical surface. In particular we focus on patterns of
chiral helices, achiral rings or vertical lamellae, with the
constraint of global electroneutrality. We study the dependence of
the patterns' size and pitch angle on the radius of the cylinder and
salt concentration. We obtain a phase diagram by using numerical and
analytic techniques. For pure Coulomb interactions, we find a ring
phase for small radii and a chiral helical phase for large radii. At
a critical salt concentration, the characteristic domain size
diverges, resulting in macroscopic phase segregation of the
components and restoring chiral symmetry. We discuss possible
consequences and generalizations of our model.
\end{abstract}

\pacs{05.70.Np, 41.20.Cv, 64.60.Cn, 82.70.Uv}

\maketitle

%%%%%%%%%%%%%%%%%%%%%%%%%%%%%%%%%%%%%%%%% LETTER %%%%%%%%%%%%%%%%%%%%%%%%%%%%%%%%%%%%%%%%%%%%

\begin{figure}[t]
\centering\includegraphics[width=0.4\textwidth]{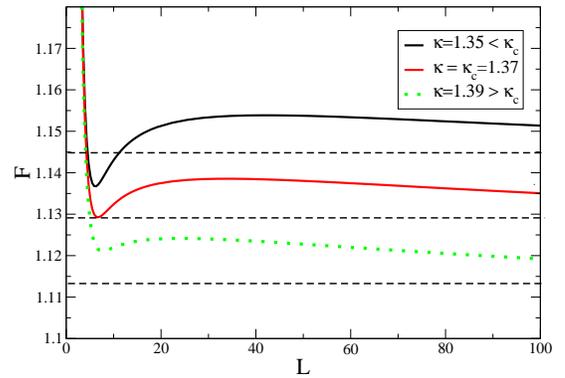} \caption{By
increasing the screening length above a critical value $\kappa_{c}$,
the free energy global minimum jumps from a finite $L$ to
$L=\infty$. In this plot $R=1000$ and  $\kappa_{c}=1.37\pm0.05$}
\label{fvsL}
\end{figure}

The ubiquitous role electrostatics play in the organization of
biomolecular assemblies is a main pillar of modern biophysics (see,
for instance, \cite{levin2} and references therein). The co-assembly
of oppositely charged molecules \cite{artzner} often occurs over
cylindrical structures with directional dependent functionalities,
such as in fibers or filamentous viruses \cite{suzuki,artzner} for
signal transmission or cellular structure mobility \cite{lehnert}.
Other important cationic-anionic cylindrical co-assemblies include
actin-binding proteins structures \cite{thomas}, cylindrical
micelles \cite{kaler}, and peptide-amphiphile fibers
\cite{stupp2,stupp3}. A considerable aspect of these systems is the
presence of surface charge heterogeneities or domains \cite{paco,
sharon1}, which arises from the competition of electrostatic forces
with segregating forces such as steric, van der Waals, and hydrogen
bonding interactions. These surface patterns have been shown to be
important for the stability and functionality
of the aggregates \cite{artzner,tombolato,israel,yuri}.\\
\indent A key feature in many biological processes is pattern
recognition and specificity within isomeric aggregates \cite{Qi}.
Biomolecules exploit repetitive patterning over surfaces, which
often breaks some underlying symmetry, in order to increase their
functionality \cite{vereb}. In this context, chiral symmetry (that
is the absence of improper rotation axis) is notably interesting:
although its role in chemistry and biology is widely recognized
\cite{artzner, pickett, Harris, kornyshev}, its origin in biology is
the subject of much debate \cite{reinhoudt}. Many systems are chiral
due to the chirality of its components \cite{reinhoudt,malinin}. For
instance, colloidal solutions of rod-like viruses provide an
intriguing example of chiral structures in liquid crystalline phases
\cite{grelet}. Nevertheless, understanding whether electrostatic
interactions alone are capable of producing chiral systems would
shed further light on the issue.\\
\indent From a theoretical vantage point, the electrostatic
patterning of a system of charges on cylindrical surfaces is however
relatively unexplored. Certainly, the effects of long-range
electrostatic forces have been widely studied for planar two
dimensional systems \cite{low}; also the behavior of short-range
interactions over cylindrical geometries has been addressed, such as
the Ising model on the cylinder \cite{cardy}.  We analyze here an
intermediate case where charges are confined over a cylindrical
surface and interact via long-range forces. A further issue we
consider is whether spherically symmetric electrostatic interactions
are capable to break translational, rotational or chiral isometries
of the cylinder. Recently, there has been interest to study
crystalline systems over constrained geometries such as the surface
of spheres, cylinders, and tori \cite{bowick,khan,endo}. The
generalization to more general curved substrates shows an
interesting rich behavior \cite{Nelson}.\\ \indent In this Letter we
provide the full phase diagram of lamellar charged patterns on a
cylinder, as a function of the cylinder radius and screening length.
We explicitly show the existence of a phase where the system
spontaneously adopts chiral configurations. We study also its
``Coulombic" origin by observing how it disappears when increasing
the screening length. Our model is a generalization of the one
introduced by F.J. Solis et al.~\cite{paco} for charge domains on
cylindrical cationic-anionic co-assemblies in aqueous solution. We
consider a 1:1 stoichiometric mixture that covers the surface
completely. The surface can generally be considered incompressible
due to the strength of hydrophobic interactions. The degree of the
incompatibility among the cationic and anionic components --
characterized by the standard Flory-Huggins parameter $\chi$ --
drives the molecules to separate and minimize the interfacial energy
between the positive and negative surface groups. While $\chi$ leads
the system to segregation, the long-range Coulomb potential favors
mixing. Molecules of like charge form homogeneous domains of average
charge density $\sigma$; boundaries between such domains acquire a
line tension $\gamma$. In the strongly segregated limit we assume
the lowest-energy configuration corresponds to domains arranged on a
periodic lattice $\{ {\mathbf \Lambda} \}$. This system can be
described by a coarse-grained two-terms free energy ${\cal F}={\cal
F}_1+{\cal F}_2$ of a Wigner-Seitz cell, with the line tension term
${\cal F}_1$ opposing the surface charge density term ${\cal F}_2$
\cite{paco}:
\begin{equation}
{\cal F}_1= \gamma l \, , \quad {\cal F}_2=\frac{1}{2} \int
\!\!\!\int' d^2 {\mathbf x} \, d^2 {\mathbf y} \, \sigma( {\mathbf
x} ) U({\mathbf x} -{\mathbf y} ) \sigma({\mathbf y} ) \, ,
% \, dz \sigma(x) \int_{cell}d\eta\hspace*{1mm}\rho(\xi)
%\rho(\mathbf{\Lambda}+\eta)V(\mathbf{\Lambda}+\eta-\xi) \, ,
\label{freeEnergy}
\end{equation}
where $l$ is the interfacial length between the domains,
$\sigma({\mathbf x})$ is the local charge density, and $\int\!\!\!
\int'$ represents an integration over the whole surface and the
other restricted to a single Wigner-Seitz unit cell. In real systems
the electrostatic potential is attenuated by free ions in the
solvent and we adopt a screened Coulomb potential:
\begin{equation}
U( {\mathbf x})= \frac{1}{4 \pi \varepsilon} \frac{e^{-\kappa |
{\mathbf x}|}}{| {\mathbf x}|} \, , \label{potential}
\end{equation}
where $\varepsilon$ is the dielectric constant, and $\kappa$ is the
inverse Debye-H\"{u}ckle length.
\begin{figure}[t]
\centering
\includegraphics[width=.4\textwidth]{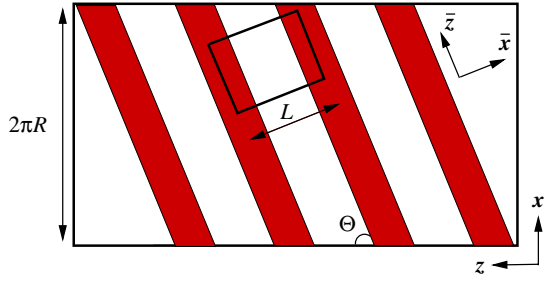}
\caption{Schematic view of an unwrapped cylinder with a generic
helical pattern, showing the notation used in the text.}\label{cyl}
\end{figure}
The latter takes effectively into account possible salt
concentration in the solution, and we will use it here also as a
tunable parameter to control the long-range/short-range nature of
the interactions. The two terms in the free energy ${\cal F}$
generate a characteristic length $L_{0}=\sqrt{\gamma
\varepsilon}/\sigma$, and it is therefore convenient to adopt
dimensionless units. All lengths $x$ are measured in units of
$L_{0}$, $\tilde{x}=x/L_0$, $\tilde{l}=l/L_0$, the inverse screening
length becomes $\tilde{\kappa}=\kappa L_0$, the charge density
$\tilde{\sigma}=\sigma/L_0^2$, and the free energy is measured in
units of ${\cal F}_0=\gamma L_0$, i.e. $\tilde{{\cal F}}={\cal
F}/{\cal F}_0$. For the sake of compactness in the rest of this
Letter we assume that the name of a variable stands for its
dimensionless counterpart, and we drop all the tilde symbols. We
consider the free energy density $F={\cal F}/A_c$, where $A_c$ is
the area of the Wigner-Seitz cell. A self-consistent minimization of
the free energy $F$ for a planar system gives lamellar patterns at
1:1 stoichiometric ratio \cite{paco}, and it has been confirmed by
direct numerical simulations \cite{sharon}. Henceforth we
concentrate on the behavior of such lamellar patterns and their
critical properties on the cylinder. In that case, we have $F_1=2/L$
where $L$ is the distance between two neighboring lamellae
\cite{paco}. The double integral in $F_{2}$ can be computed first by
``unwrapping" the surface of the cylinder of radius $R$ onto an
infinite set of parallel stripes of size $2\pi R$ on the plane, and
then by summing all the pairwise electrostatic interactions over the
planar periodic lattice $\{\mathbf{\Lambda}\}$ of the lamellae, by
Fourier decomposition method. To simplify the calculations we
adopted a square lattice oriented along the lamellae (see Fig.
\ref{cyl}). Formally:
\begin{equation}
F_{2}= \frac{1}{2}\sum_{\mathbf{\Lambda}} \int_{cell}d\xi
\int_{cell}d\eta\hspace*{1mm}\sigma(\xi)
\sigma(\mathbf{\Lambda}+\eta)V(\mathbf{\Lambda}+\eta-\xi) \, ,
\label{s2}
\end{equation}
where
\begin{equation}
V({\mathbf{p}})=\frac{e^{-\kappa D}}{4 \pi D}\, \theta(2 \pi
R-|p_x|)\, ,
\end{equation}
with $D \equiv \sqrt{4R^{2}\sin^{2} \left( p_x/(2R)\right)
+p_z^{2}}$ is the projected distance of $\mathbf{p} \equiv(p_x,p_z)$
on the $\{ x,z \}$ plane, and $\theta(x)$ is the step function.
\begin{figure}[t]
\centering \includegraphics[width=8.5cm]{phase}
\put(-155,30){\makebox{\includegraphics[width=.5cm]{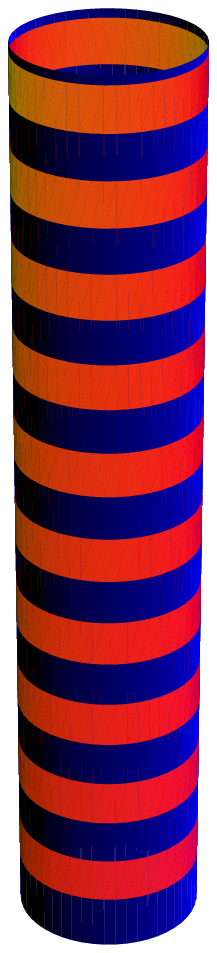}}}
\put(-70,40){\makebox{\includegraphics[width=.5cm]{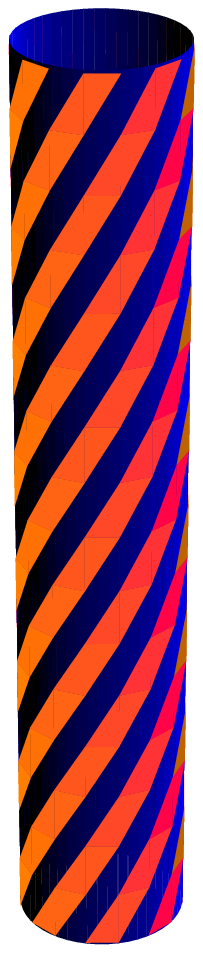}}}
\put(-20,125){\makebox{\includegraphics[width=0.4cm]{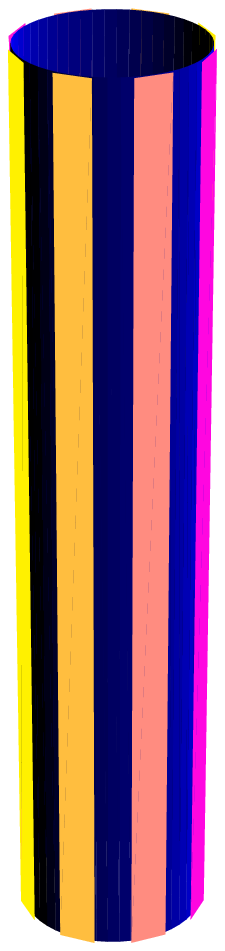}}}
\put(-180,90){\makebox{\includegraphics[width=0.45cm]{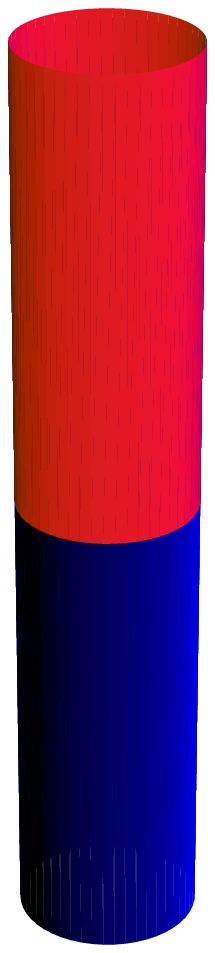}}}
\caption{ Phase diagram of our model. Region $\textbf{I}$ is the
ring phase; Region $\textbf{II}$ is the helical phase; Region
$\textbf{III}$ is the axial phase; the region at the top is the
fully segregated phase. The solid lines are discontinuous
transitions and the dashed line is a continuous crossover.
}\label{phase}
\end{figure}
By introducing the reciprocal lattice $\{\mathbf{Q} \}$, defined by
$\mathbf{Q}\cdot\mathbf{\Lambda}=2\pi m$, $m \in \mathbb{Z}$, and by
using the Poisson summation formula
 we have:
\begin{equation}
F_{2}=\frac{1}{2
A_c}\sum\limits_{\mathbf{Q}}\hat{\sigma}(\mathbf{Q})\hat{\sigma}(-\mathbf{Q})\hat{V}(\mathbf{Q})
\, , \label{s22}
\end{equation}
where $\hat{\sigma}(\mathbf{Q})$ and $\hat{V}(\mathbf{Q})$ are the
Fourier transforms of $\sigma$ and $V$, respectively. They are:
\begin{eqnarray}
\hat{V}(\mathbf{Q})&=& R\,  I_{q_{x} R} \left( \xi \right)
K_{q_{x} R} \left( \xi\right),  \label{vhat}\\
\hat{\sigma}(\mathbf{Q})&=&4\frac{\sin\left( {\bar{q}}_{x}/4 \right)
}{{\bar{q}}_{x}}\delta({\bar{q}}_{z}) \, , \label{density}
\end{eqnarray}
where $\xi \equiv R \sqrt{ \kappa^2+q_z^2}$ and $I_{\nu}$ and
$K_{\nu}$ are modified Bessel functions of order $\nu$ of the first
kind and second kind, respectively. For simplicity we used the
twofold notation $\mathbf{Q}=({\bar{q}}_x,{\bar{q}}_z)$ in the
reference system parallel to the lamellar domains, and
$\mathbf{Q}=(q_x,q_z)$ in the reference system parallel to the axis
of the cylinder (see Fig.~\ref{cyl}). Let $\Theta$ be the pitch
angle between the axis of the cylinder $z$ and the direction of the
lamellar stripes $\bar{z}$. Any lattice on the surface of a cylinder
must be commensurate with its circumference, that is it must be
periodic in the $x$ direction with period $2\pi R$. The
commensurability constraint in the reciprocal space read
$\bar{q}_x=2 \pi m/L$ or equivalently $q_x=2\pi m \cos\Theta / L$
and $q_z=2\pi m\sin\Theta / L$, $m \in \mathbb{Z}$. Moreover $q_x
R= m n$ must hold, where $n$ is the number of lamellar stripes per pitch.\\
We numerically minimized the free energy density $F$ with respect to
all possible lamellar lattices, i.e. with respect to the spacing $L$
and the pitch angle $\Theta$ for several values of $R$ and $\kappa$.
Our results are collected in the phase diagram in Fig.~\ref{phase}.
There are four regions: a phase $\textbf{I}$ characterized by ring
patterns, a chiral phase $\textbf{II}$ characterized by helical
patterns, a phase $\textbf{III}$ characterized by vertical lamellar
patterns, and a fully segregated
phase.\\
\begin{figure}[t]
\centering \includegraphics[width=.4\textwidth]{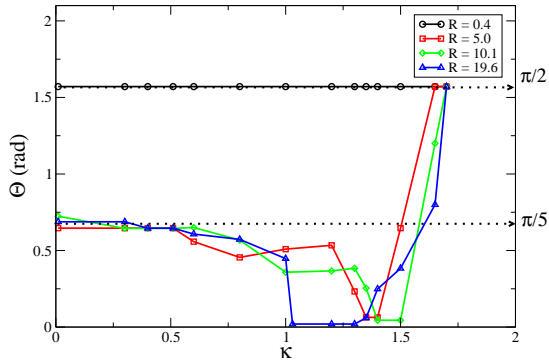}
\caption{Evolution of the pitch angle $\Theta$ of helical lamellae
with the Debye screening inverse-length $\kappa$, at different
cylinder radii $R$.} \label{salty}
\end{figure}
\indent Phase $\textbf{I}$ is limited to fibers with circumference
smaller than the domain size. This is natural when considering the
limit $R \ll L$ where $\cos \Theta \simeq nL/2\pi R \to 1$, i.e.
$\Theta \simeq \pi/2$, which is precisely the observed ring phase.\\
\indent Phase $\textbf{II}$ is the primary result of this work. It
proves the existence of chiral configurations (both right-handed and
left-handed) due to electrostatic long-range interactions. A
surprising feature of this phase is that at $\kappa=0$
%and sufficiently large $R$
there is a preferred pitch angle $\Theta^{*} \simeq \pi/5$ (see
Fig.~\ref{salty}). The long-range nature of electrostatic
interaction may be the reason of such a preference, since by
increasing $\kappa>0$ the pitch angle decreases continuously down to
the vertical lamellar phase $\textbf{III}$. Our numerical findings
can be further justified on a theoretical ground by looking at the
large-$R$ asymptotic expansion of the free energy. In the expansion
$q_x$ is kept fixed while the density $\sigma(\mathbf{x})$ is not
dependent on $R$ since it is defined locally over a unit cell. For
the potential $\hat{V}(\mathbf{Q})$ we obtain (up to $O(1/R^4)$
terms):
\begin{equation}
\hat{V}(\mathbf{Q})\sim\frac{1}{2\sqrt{Q^2+\kappa^2}}-
\frac{\left(4Q^2-5q^{2}_{z}-\kappa^2 \right)\left(q^{2}_{z}+\kappa^2
\right)}{16 \left( Q^{2}+\kappa^2 \right)^{7/2}}\frac{1}{R^2}
\label{asympt}
\end{equation}
where $Q\equiv\sqrt{q_z^2+q_x^2}$. The leading term is the planar
limit studied in \cite{sharon}. It is straightforward to show that
the minimum of the free energy $F$ at $\kappa=0$ with respect to the
pitch angle is where $d \hat{V}(\mathbf{Q})/d \Theta=0$. From
eq.~(\ref{asympt}) at $\kappa=0$ we find $\Theta^{*}=\arccos
\sqrt{3/5}\simeq 0.68$ rad which is close to the numerical value
$\pi/5\simeq 0.63$ rad. For $\kappa>0$ the first-order correction to
$\Theta^{*}$ is negative and scale as $\kappa^2$. This is in
agreement with the numerical behavior depicted in Fig.~\ref{salty},
that the pitch angle initially decreases when increasing $\kappa$.
From eq.~(\ref{asympt}) we obtain also the value
$\kappa_0=2\sqrt{2/3}\pi/L$ at which the pitch angle vanishes, i.e.
where phase $\textbf{II}$ merges continuously into $\textbf{III}$
and the helical patterns evolve into vertical stripes. We find
$L=5.8\pm0.2$ at large-$R$ (see Fig.~\ref{fvsL}), and therefore
$\kappa_0=0.9\pm0.2$. The nature of phase $\textbf{II}$ can further
be seen in the discontinuous transition to the achiral phase
$\textbf{I}$. Assuming the separation distance $L$ is not strongly
dependent on $R$, there is a critical ratio $L/R$ such that, as $R$
decreases, $\Theta$ discontinuously jumps from $\Theta^{*}$ to
$\pi/2$ (see Fig.~\ref{salty}).

All the three phases we considered so far are bound at large
$\kappa$ by a fully-segregated phase. Such a salt-induced phase
transition is a first-order phase transition as the free energy
jumps from a finite cell size to an infinite cell size (see
Fig.~\ref{fvsL}). Let $\kappa_c(R)$ be the critical value of
$\kappa$ at which this transition occurs.
In Fig.~\ref{salty} the transition is at the jump of the pitch to
$\pi/2$, near the ``melting'' transition point $\kappa \sim
\kappa_c$. At large-$R$ we find $\kappa_c=1.6\pm0.1$. The melting
transition can be interpreted by a scaling argument. At large-$L$
the free-energy scales as $F\approx1/L+1/\kappa$. If $\kappa$ is
small then the global minimum is at a finite value of $L$, that is
the size of the Wigner cell (see continuous black line in
Fig.~\ref{fvsL}). If $\kappa$ is large then the minimum jumps to
infinity, that is to a macroscopically segregated system (see dotted
green line in Fig.~\ref{fvsL}).

We note that the phase diagram for the cylinder is richer than its
planar counterpart. In fact,  due to the commensurability
constraints of the cylindrical surface new lamellar pattern
configurations with different chiral angles arise for different
radii $R$ and lamellar spacing $L$. The theoretical prediction for
the cylindrical geometry in the large radius limit agrees well with
the result for planar geometry. However a striking distinction from
the planar case is the appearance of a preferred chiral angle of
about $\Theta^{*}=\arccos \sqrt{3/5} \approx \pi/5$, which is
numerically found to be conserved over a wide range of radii. It is
an amusing coincidence this angle corresponds to the one found in
recent studies of capsid-protein arrangements on cylindrical viruses
\cite{grelet} as well as in biomimetic self-organized nanotubes
\cite{artzner}. However Coulomb interaction is not in principle the
driving force of the assemblies considered in \cite{grelet,artzner}.
Therefore an interesting question is whether such a pitch angle is
universal, and what is its origin. We attempt an answer by
considering the large-$R$ asymptotic expansion for higher-order
decaying isotropic interactions, $V_{3}\sim 1/D^3$. We find (up to
$O(1/R^4)$ terms: $ \hat{V}_{3}(\mathbf{Q})\sim -\pi Q+ \pi
q^{2}_{z}/(8 Q^5 R^2)$, whose stationary points are only at
$\Theta=0$ and $\pi/2$ (this statement holds true also for higher
multipolar terms). Although this argument cannot be considered as
proof, it shows that the universality of the pitch angle is
unlikely. However it also shows that the $\sim \pi/5$ angle seems to
arise in conjunction with the Coulomb potential. As we have seen,
the exponential screening term actually softens this effect through
the reduction of the pitch angle, but it does not suppress it
entirely (for small values of $\kappa$). From this point of view,
the chiral symmetry we observe originates only from the interplay
between the $1/D$ behavior of the Coulomb potential and the
cylindrical geometry. Such an effect is independent from the
specific pattern $\sigma$ and therefore we believe it might emerge
also in more general
electrostatic systems over a cylinder.\\
\indent The symmetry group of a decorated cylinder falls into one of
9 classes, each with a distinct closed subgroup of the full cylinder
group ${\mathbf S}{\mathbf O}(2)\times\mathbb{R}$ \cite{golubitsky}.
The four regions in our diagram correspond to the four classes
characterized by one-dimensional closed subgroups (the rest being
zero- or two-dimensional). Namely, up to conjugacy and scaling, they
are \cite{golubitsky}: pure rotations ${\mathbf S}{\mathbf
O}(2)\times\mathbb{I}$ (fully segregated phase), continuous
rotations and discrete translations ${\mathbf S}{\mathbf
O}(2)\times\mathbb{Z}$ (phase {\bf I}), discrete rotations and
continuous translations ${\mathbf Z}_k\times\mathbb{R}$ (phase {\bf
III}), and corkscrew symmetries ${\mathbf Z}_k\times{\mathbf L}$
(phase {\bf II}). ${\mathbf Z}_k$ is the subgroup of rotations of
the cylinder through angles that are multiples of $2\pi/k$, and
${\mathbf L}=\{(t,t)\in {\mathbf S}{\mathbf O}(2)\times \mathbb{R} :
t \in \mathbb{R} \}$. Therefore the phase transitions we observe
correspond to breaking
qualitatively different symmetry generators.\\
\indent We expect these phases to be relevant for systems at the
nanoscale, as it can be easily seen by an order-of-magnitude
analysis of $L_0$. It is through these length scales we believe the
functionality of the patterns might manifest, such as the induced
chirality in our system. The advantage of the nanoscale, besides
possible practical applications, is that electrostatic interactions
are still important under normal dielectric conditions. Even in
aqueous systems where charge screening occurs, Coulomb interactions
still play a vital role through counterions and metallic-ion
coordination \cite{lehn}. Finally, we do not explore in this Letter
the possibility of a critical point where phase {\bf I}, {\bf II} or
the fully segregated
phase meet, and we postpone it to future studies.\\
\indent The authors acknowledge Dr. Y. Velichko for useful
discussions. This work was supported in part by the NSF (Grant No.
DMR-0414446) and by a graduate fellowship of the DOE (Grant No.
DE-FG02-97ER25308 for KLK).

\end{document}